\documentclass[pra,twocolumn,showpacs]{revtex4}
\usepackage{epsfig}
\usepackage{amsmath}

\begin{document}

\title{Qubit coupled with an effective negative-absolute-temperature bath in
off-resonant collision model}
\author{Wei-Bin Yan}
\affiliation{College of Physics and Engineering, Qufu Normal University, Qufu, 273165,
China}
\author{Zhong-Xiao Man}
\affiliation{College of Physics and Engineering, Qufu Normal University, Qufu, 273165,
China}
\author{Ying-Jie Zhang}
\affiliation{College of Physics and Engineering, Qufu Normal University, Qufu, 273165,
China}
\author{Yun-Jie Xia}
\affiliation{College of Physics and Engineering, Qufu Normal University, Qufu, 273165,
China}

\begin{abstract}
Quantum collision model provides a promising tool for investigating
system-bath dynamics. Most of the studies on quantum collision models work
in the resonant regime. In quantum dynamics, the off-resonant interaction
often brings in exciting effects. It is thereby attractive to investigate
quantum collision models in the off-resonant regime. On the other hand, a
bath with a negative absolute temperature is anticipated to be instrumental
in developing thermal devices. The design of an effective bath with negative
absolute temperature coupled to a qubit is significant for developing such
thermal devices. We establish an effective negative-absolute-temperature
bath coupled to a qubit with a quantum collision model in a far-off-resonant
regime. We conduct a detailed and systematic investigation on the
off-resonant collision model. There is an additional constraint on the
collision duration resulting from the far-off resonant collision. The
dynamics of the collision model in the far-off-resonant regime are different
from the one beyond the far-off-resonant regime. Numerical simulations
confirm the validity of the proposed approach.
\end{abstract}

\maketitle

\section{introduction}

Quantum collision model is a powerful and effective tool for investigating
the dynamics of the system-bath scheme \cite%
{Lorenzo2015,Lorenzo20151,Pezzutto_2016,Strasberg2017,Cusumano2018,Seah2019,Rodrigues2019,Barra2019,Molitor2020,Taranto2020,Campbell_2021,2022Quantum,zhangq2023,zhangq2023_1}%
. It first appeared in 1963 \cite{Rau1963} and has gained growing use in the
last few years. In quantum collision model, the bath is represented by an
extensive collection of subunits (ancillas). The open system is coupled to
the bath by the system-ancilla short collisions, i.e., short unitary
interactions. The collisions are performed through a pairwise sequence. In
most quantum collision models, the system resonantly interacts with the
ancillas. In investigating the dynamics of a quantum system, it is common to
encounter a situation in which quantum objects off-resonantly interact with
each other. These off-resonant interactions often bring fresh effects \cite%
{James_2007} compared to the resonant interaction. For example, a two-level
system far-off-resonantly driven by an external weak Laser would gain slight
shifts of its levels, known as the \textit{A. C. Stark shift}. An atom with
an appropriate structure interacting with light can induce a large nonlinear
effect when light off-resonantly drives the specific level transition \cite%
{1996Giant,2006NatPh}. In quantum technology, a widely used approach to
couple the decoupled levels of interest is to introduce the intermediate
auxiliary levels working in the off-resonant regime, as shown in the
examples (in the vast number of instances) in Refs. \cite%
{Hartmann2007,Zheng2000,Campbell2020,Zhang2024}. Therefore, it is reasonable
that quantum collision model in the off-resonant regime, i.e., the system
off-resonantly colliding with the ancillas, may bring in fresh effects
compared to the resonant case. A systematic investigation of the quantum
collision model in the off-resonant regime would inject more vitality into
the field of quantum collision model.

In recent years, negative absolute temperature has caused intensive
attention. Onsager originally conceived the physical idea of the negative
absolute temperature in the statistical investigation of the point vortices
\cite{Onsager}, which has been demonstrated in recent experiments with the
two-dimensional quantum superfluid \cite{Johnstone,Gauthier}. Subsequently,
Purcell and Pound observed the negative absolute temperatures in the nuclear
spin systems with LiF crystal \cite{Purcell}. Ramsey theoretically
investigated the thermodynamic and statistical mechanical implications of
such negative absolute temperatures \cite{Ramsey}. While there is some
confusion regarding the equilibrium of negative absolute temperature related
to the definition of entropy \cite{2014Consistent,CALABRESE20192153},
negative absolute temperatures are now widely accepted in the scientific
community and appear consistent with experimental observations \cite%
{Frenkel2015,BUONSANTE2016414,PUGLISI20171,Cerino2015,Abraham2017,BALDOVIN20211,Onorato2022,Baudin2023,Muni2023}%
. The negative absolute temperature bath is expected to develop thermal
devices such as Carnot engines \cite%
{Geusic1967,Landsberg_1977,Nakagomi_1980,Dunning-Davies_1976,Dunning1978,Landsberg_1980,Xi_2017,Assis2019,Nettersheim2022,maity2023,Maity2024,desousa2024}
and refrigerators \cite{DAMAS2023129038}, in which the devices involved in
the negative absolute temperature bath would perform better than their
traditional counterparts based on positive temperatures. In this context,
for most proposals, the negative temperature bath was assumed to exist
already or was realized by external driving or work. Notably, the authors in
Ref. \cite{Bera2024} proposed an approach to realize a synthetic negative
temperature bath coupled to a three-level atom. The temperature of the
synthetic bath could be an arbitrary value in the range from $-\infty $ to $%
+\infty $. It opens up an issue on how to realize a negative temperature
bath coupled to a qubit, a unit of quantum information widely used in
constructing quantum thermal devices.

In this paper, we propose to realize an effective negative temperature bath
coupled to a qubit in the context of quantum collision model in the
off-resonant regime. The qubit is effectively realized by adiabatically
eliminating the highest level of a $\Lambda $-type qutrit when the ancillas
of the effective bath far-off resonantly collide with the qutrit. Under
certain conditions, based on the dynamics of the proposed collision scheme,
a Gorini-Kossakowski-Sudarshan-Lindblad master equation, which represents
the dynamics of a qubit coupled to an effective bath with adjustable
temperature, is found by performing appropriate approximations. The validity
of the master equation is confirmed by numerical simulations, which is
necessary to ensure the rigor of this work because we have not found any
precedent to handle quantum collision models in our way. We find that an
additional constraint on the collision duration is impressed since the lower
bound of the coarse-graining time stems from the rapid oscillation dynamics
of the far-off-resonant collisions. Beyond the far-off-resonant-collision
case, one can not realize the effective qubit-bath coupling because it is difficult to
neglect the dynamics of the highest level of the qutrit. In
this case, the dynamics are similar to those outlined in Ref. \cite{Bera2024}%
.

\section{Collision Model}

\begin{figure}[t]
\centering
\includegraphics[width=6cm]{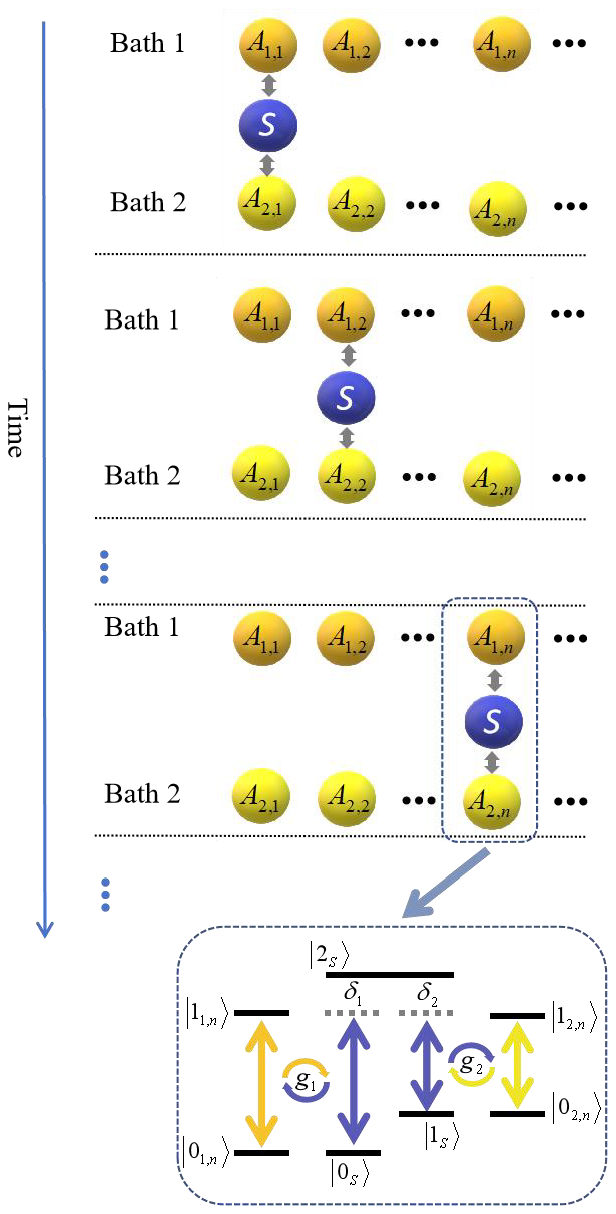}\newline
\caption{Collision model: the system S composed of a three-level qutrit
undergoes sequence successive collisions with the ancilla qubits in bath 1
and bath 2.}
\label{MODEL}
\end{figure}

The collision model under consideration is sketched in Fig. \ref{MODEL}. The
open system S consists of a $\Lambda $-type qutrit. The three levels of S
are labeled by $\left\vert 0_{S}\right\rangle $, $\left\vert
1_{S}\right\rangle $, and $\left\vert 2_{S}\right\rangle $, with the
corresponding level frequencies denoted by $\omega _{S,0}$, $\omega _{S,1}$,
and $\omega _{S,2}$, respectively. There are two baths, either of which
contains many identical ancillas, i.e., noninteracting qubits, coupled to
the open system via a sequence of system-ancilla collisions. The $n$-th ($%
n=1,2,3...$) ancilla of bath $m$ ($m=1,2$) is labeled by $A_{m,n}$. The
upper and lower levels of $A_{m,n}$ are denoted by $\left\vert
1_{A_{m,n}}\right\rangle $ and $\left\vert 0_{A_{m,n}}\right\rangle $,
respectively. The frequency corresponding to the upper (lower) level of bath
$m^{\prime }$s identical qubits is labeled by $\omega _{A_{m,1}}$ ($\omega
_{A_{m,0}}$). We focus on the collisional memoryless model. Each ancilla
collides with the system S only once. The $n$-th collision step is realized
by the short interaction of S with $A_{1,n}$ and $A_{2,n}$. The system level
transitions $\left\vert 0_{S}\right\rangle \leftrightarrow \left\vert
2_{S}\right\rangle $ and $\left\vert 1_{S}\right\rangle \leftrightarrow
\left\vert 2_{S}\right\rangle $ are coupled to the ancilla level transitions
$\left\vert 1_{A_{1,n}}\right\rangle \leftrightarrow \left\vert
0_{A_{1,n}}\right\rangle $ and $\left\vert 1_{A_{2,n}}\right\rangle
\leftrightarrow \left\vert 0_{A_{2,n}}\right\rangle $, respectively. In the
interaction picture, the Hamiltonian representing the $n$-th collision step
reads%
\begin{equation}
H_{n}=g_{1}\sigma _{A_{1,n}}^{10}\sigma _{S}^{02}e^{-i\delta
_{1}t}+g_{2}\sigma _{A_{2,n}}^{10}\sigma _{S}^{12}e^{-i\delta _{2}t}+h.c.,
\label{H_INT}
\end{equation}%
where $\sigma _{M}^{kk^{\prime }}=\left\vert k_{M}\right\rangle \left\langle
k_{M}^{\prime }\right\vert $ is $M$'s level transition and population
operators with $M\in \{A_{1,n},A_{2,n},S\}$. The interaction strength
between $A_{m,n}$ and the system is denoted by $g_{m}$, which remains
unchanged with the collision step number $n$. The detunings $\delta _{1}$
and $\delta _{2}$ are represented by $\delta _{1}=\omega _{S,20}-\omega
_{A_{1}}$ and $\delta _{2}=\omega _{S,21}-\omega _{A_{2}}$, respectively, as
shown in Fig. \ref{MODEL}. Here $\omega _{S,kk^{\prime }}=\omega
_{S,k}-\omega _{S,k^{\prime }}$ denotes the system level transition
frequency between the levels $\left\vert k_{S}\right\rangle $ and $%
\left\vert k_{S}\right\rangle $, and $\omega _{A_{m}}=\omega
_{A_{m,1}}-\omega _{A_{m,0}}$ is the level transition frequency of the qubit
in bath $m$. We have set the reduced Plank constant $\hbar $ as a unit,
i.e., $\hbar =1$. The first (second) term in Hamiltonian (\ref{H_INT})
represents the off-resonant interaction between $A_{1,n}$ ($A_{2,n}$) and S
when $\delta _{1}$ ($\delta _{2}$) is nonzero. In the far-off-resonant case,
i.e., the large-detuning case represented as $\delta _{m}\gg g_{m}$, one can
perform the standard adiabatic elimination \cite{James_2007} and obtain the
effective Hamiltonian as%
\begin{eqnarray}
H_{eff,n} &=&-\alpha _{11}\sigma _{A_{1,n}}^{11}\sigma _{S}^{00}-\alpha
_{22}\sigma _{A_{2,n}}^{11}\sigma _{S}^{11}  \notag \\
&&-\alpha _{12}\sigma _{A_{1,n}}^{01}\sigma _{A_{2,n}}^{10}\sigma
_{S}^{10}e^{i(\delta _{1}-\delta _{2})t}+h.c.,  \label{H_EFF}
\end{eqnarray}%
where $\alpha _{mn}=\frac{g_{m}g_{n}}{2}(\frac{1}{\delta _{m}}+\frac{1}{%
\delta _{n}})$. The vital condition $\left\vert \delta _{1}-\delta
_{2}\right\vert \ll \alpha _{12}$ should be satisfied so that the dynamics
governed by the second line of the effective Hamiltonian (\ref{H_EFF}) play
a significant role and are not ignored. The effective Hamiltonian eliminates
the system's highest level $\left\vert 2_{S}\right\rangle $, which is
referred to as a virtual level in the subsequent analysis. Either of the
terms in the first line of the effective Hamiltonian can be understood by a
sequence of two virtual procedures. Taking the first term as an example, in
the first virtual procedure, $A_{1,n} $ makes the transition $\left\vert
1_{A_{1,n}}\right\rangle \rightarrow \left\vert 0_{A_{1,n}}\right\rangle $,
meanwhile, S jumps from the level $\left\vert 0_{S}\right\rangle $ to the
virtual level. In the second virtual procedure, $A_{1,n}$ flips back to
level $\left\vert 1_{A_{1,n}}\right\rangle $; meanwhile, S flips back to
level $\left\vert 0_{S}\right\rangle $. The second line can be understood from the fact that
S jumps between the levels $\left\vert 0_{S}\right\rangle $ and $\left\vert
1_{S}\right\rangle $ via the virtual level; meanwhile, $A_{1,n}$ and $%
A_{2,n} $ accomplish their corresponding level transitions.

The terms in the first line of the effective Hamiltonian result in slight
shifts of the corresponding levels. To understand this level-shift effect,
one can consult the analysis represented by Ref. \cite{PNeumeie2013}, in
which similar interactions are investigated. The effective Hamiltonian
implies that the system S is an effective qubit with the levels $\left\vert
0_{S}\right\rangle $ and $\left\vert 1_{S}\right\rangle $. The two ancillas
cooperatively drive the common effective-qubit-level transition in each
collision step. Here, we take the condition $\delta _{1}=\delta _{2}=\delta $
so that the energy absorbed (or released) by the effective-qubit-level
transition equals the energy released (or absorbed) by the cooperative
ancilla-level transitions.

\begin{figure}[t]
\centering
\includegraphics[width=8cm]{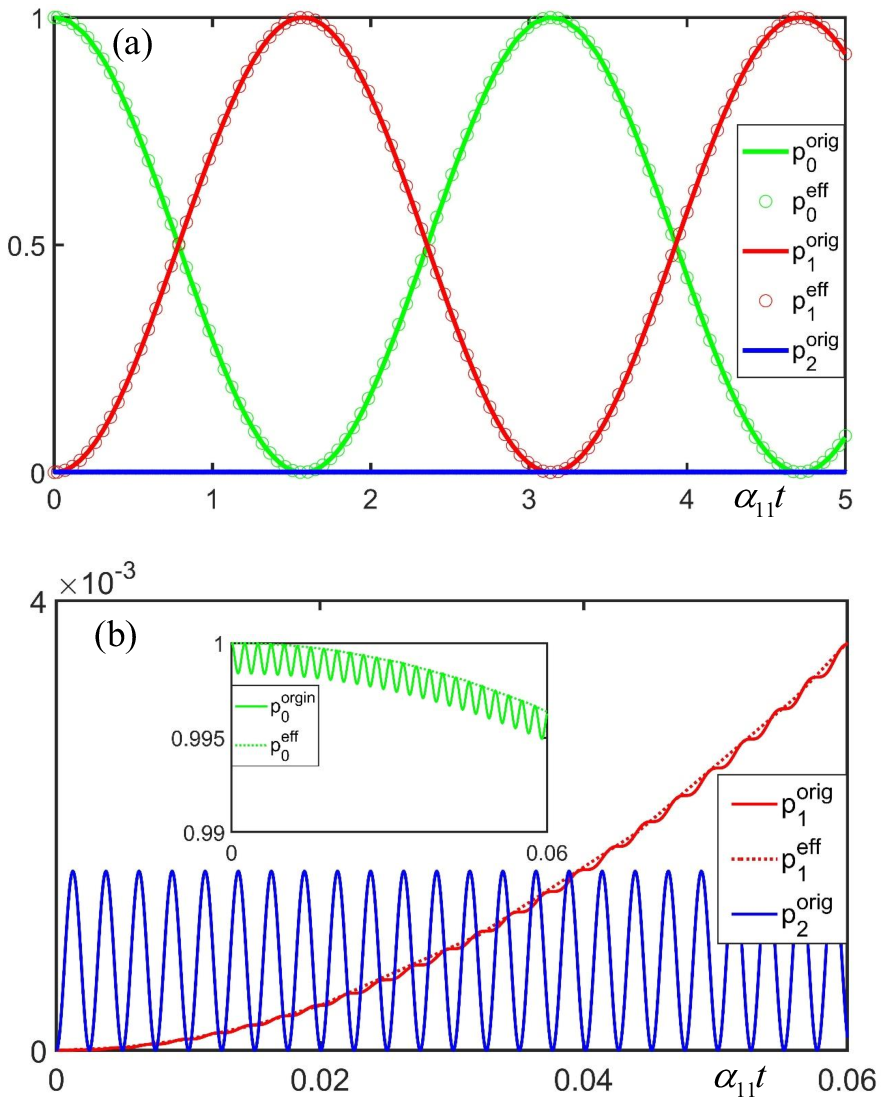}\newline
\caption{Numerical simulation of the populations against the time governed
by Hamiltonian (\protect\ref{H_INT}) and (\protect\ref{H_EFF}). The initial
state is set as $\left\vert 1_{A_{1,n}},0_{A_{2,n}},0_{S}\right\rangle $.
(a) shows the evolutions of the populations from $\protect\alpha _{11}t=0$
to $\protect\alpha _{11}t=5$, and (b) shows the evolutions from $\protect%
\alpha _{11}t=0$ to $\protect\alpha _{11}t=0.006$. In (b), in order to
observe the slight oscillations of evolutions governed by Hamiltonian (%
\protect\ref{H_INT}), the populations of the state $\left\vert
1_{S}\right\rangle $ is represented as the insert panel because the
difference between the population of $\left\vert 0_{S}\right\rangle $ and
the population of $\left\vert 1_{S}\right\rangle $ is much larger than the
amplitudes of the oscillations. The green, red, and blue solid lines denote $%
p_{0}^{orig}$,\ $p_{1}^{orig}$, and $p_{2}^{orig}$, respectively. In (a), $%
p_{0}^{orig}$ and $p_{1}^{orig}$ are represented by green and red hollow
circles, respectively. While, in (b), $p_{0}^{orig}$ and $p_{1}^{orig}$ are
represented by green and red dotted lines, respectively. The parameters are $%
g_{2}=g_{1}$ and $\protect\delta =50g_{1}$.}
\label{VERIFY}
\end{figure}

The legitimacy of the adiabatic elimination on the level $\left\vert
2_{S}\right\rangle $ is confirmed in Fig. \ref{VERIFY}, which numerically
simulates the dynamics governed by the original Hamiltonian (\ref{H_INT})
and effective Hamiltonian (\ref{H_EFF}) in the far-off-resonant case. The
symbols $p_{k}^{orig}$ and $p_{k}^{eff}$ denote the populations of $%
\left\vert k_{S}\right\rangle $ obtained by solving the original Hamiltonian
and effective Hamiltonian, respectively. In Fig. \ref{VERIFY} (a), the
evolution of $p_{0}^{eff}$ ($p_{1}^{eff}$) agrees well with the evolution of
$p_{0}^{orig}$ ($p_{1}^{orig}$). The population of the level $\left\vert
2_{S}\right\rangle $ governed by the original Hamiltonian is always near
zero. As shown in Fig. \ref{VERIFY} (b), the evolutions of the populations
governed by the original Hamiltonian show extremely slight but rapid
oscillations against time. One of the critical approximations performed in
the adiabatic elimination is that the effective Hamiltonian has considered
the coarse-grained (or time-averaged) dynamics of the original Hamiltonian.
The slight oscillations can not be visually observed in Fig. \ref{VERIFY}
(a) because the evolution time is much larger than the coarse-graining time,
and the maximum values of the populations are much larger than the
oscillation amplitudes.

\section{Continuous-Time Master Equation\label{SME}}

It will be convenient to bring in the rotating frame with respect to the
term $-\alpha _{11}\sigma _{A_{1,n}}^{11}\sigma _{S}^{00}-\alpha _{22}\sigma
_{A_{2,n}}^{11}\sigma _{S}^{11}$. In this rotating frame, the effective
Hamiltonian turns out to be%
\begin{equation}
V_{n}=-\alpha \sigma _{A_{1,n}}^{01}\sigma _{A_{2,n}}^{10}\sigma
_{S}^{10}+h.c.,  \label{H_ROT}
\end{equation}%
where we have taken $g_{1}=g_{2}=g$ and hence $\alpha _{11}=\alpha
_{22}=\alpha _{12}=\alpha $. The collision durations in all the steps are
assumed equal and denoted by $\tau $. The unitary evolution operator $%
U_{n}=e^{-iV_{n}\tau }$ represents the evolution of the n-th collision step.

Before the n-th collision step, the system and the n-th qubits of the two
baths are considered in the state $\sigma _{n-1}=\rho _{n-1}\otimes \eta
_{A_{1,n}}\otimes \eta _{A_{2,n}}$, where the density operators $\rho _{n-1}$%
, $\eta _{A_{1,n}}$, and $\eta _{A_{2,n}}$ represent the states of S, $%
A_{1,n}$, and $A_{2,n}$, respectively. Initially, the qubits in bath 1 and 2
are in the Gibbs thermal state with inverse temperatures $\beta _{1}$ and $%
\beta _{2}$, respectively, i.e.,
\begin{eqnarray}
\eta _{A_{m,n}}=\frac{\sigma _{A_{m,n}}^{11}+e^{\omega _{A_{m}}\beta
_{m}}\sigma _{A_{m,n}}^{00}}{e^{\omega _{A_{m}}\beta _{m}}+1}.  \notag
\end{eqnarray}

By performing the approximation up to the second-order of $\tau $, one can
obtain%
\begin{eqnarray}
\Delta \rho _{n} &=&Tr_{A_{1},A_{2}}(U\sigma _{n-1}U^{\dagger }-\sigma
_{n-1})  \notag \\
&\sim &Tr_{A_{1},A_{2}}(-i\tau \lbrack V_{n},\sigma _{n-1}]+\tau
^{2}(V_{n}\sigma _{n-1}V_{n}  \notag \\
&&-\frac{1}{2}[\sigma _{n-1},V_{n}^{2}]_{+}),  \label{DELTA_PHO}
\end{eqnarray}%
where the anti-commutator $[\cdot,\cdot]_{+}$ satisfies $[A,B]_{+}=AB+BA$.
Eqn. (\ref{DELTA_PHO}) denotes the stroboscopic representation of the system
dynamics with the discrete-time variables $t_{n}=n\tau $ ($n=1,2,3...$).
When the collision duration $\tau $ is much smaller than the evolution time
scale, one can consider the continuous-time limit, i.e. $\frac{\Delta \rho
_{n}}{\tau }\rightarrow \frac{d\rho }{dt}$, and obtain a continuous-time
master equation as%
\begin{eqnarray}
\frac{d\rho }{dt} &=&\Gamma e^{\omega _{S,10}\beta _{S}}(\sigma
_{S}^{01}\rho \sigma _{S}^{10}-\frac{1}{2}[\rho ,\sigma _{S}^{11}]_{+})
\notag \\
&&+\Gamma (\sigma _{S}^{10}\rho \sigma _{S}^{01}-\frac{1}{2}[\rho ,\sigma
_{S}^{00}]_{+}),  \label{ME}
\end{eqnarray}%
where $\beta _{s}=\frac{\omega _{A_{1}}\beta _{1}-\omega _{A_{2}}\beta _{2}}{%
\omega _{A_{1}}-\omega _{A_{2}}}$ and $\Gamma=\frac{R^{2}g^{2}\tau }{%
(1+e^{\omega _{A_{1}}\beta _{1}})(1+e^{-\omega _{A_{2}}\beta _{2}})}$ with $R
$ the ratio of coupling strength to the detuning, i.e., $R=\frac{g}{\delta }$%
. $R$ is dimensionless when the dimension of $g$ is the same as that of $%
\delta $. The steady-state solution of Eqn. (\ref{ME}) can be easily found as%
\begin{equation}
\rho _{ss}=\frac{\sigma _{S}^{11}}{1+e^{\omega _{S,10}\beta _{S}}}+\frac{%
e^{\omega _{S,10}\beta _{S}}\sigma_{S}^{00}}{1+e^{\omega _{S,10}\beta _{S}}},
\label{STEADY}
\end{equation}%
representing that the effective qubit is the thermal state with the inverse
temperature $\beta _{s}$. It is as if an effective bath synthesized by
integrating bath 1 and bath 2 is coupled to the effective qubit. The inverse
temperature $\beta _{s}$ of the effective bath can be tuned by the
level-transition frequency and initial Gibbs state of the ancillas. The
far-off-resonant collision model outlines an approach that couples a qubit
to an effective bath with an arbitrary temperature, including a negative
temperature.

\section{Collision Duration and Coarse-Graining Time\label{SCD}}

The short but finite collision duration is a coarse-graining time, and the
collision step is a coarse-grained procedure. We recall that the rapid
slight oscillation dynamics governed by Hamiltonian (\ref{H_INT}) has been
coarse-grained by the effective Hamiltonian (\ref{H_EFF}) under the
large-detuning condition, which naturally arises a constraint on the
collision duration. The constraint is that the collision duration should be
significantly longer than the time scale of the rapid-slight-oscillation
period. We proceed to demonstrate the validity of the continuous-time master
equation (\ref{ME}) numerically under this constraint.

One should return to the original Hamiltonian (\ref{H_INT}) to numerically
simulate the collided system's accurate dynamics. It will be convenient to
look for a new rotating frame in which the Hamiltonian is time-independent.
In the rotating frame with respect to $-\delta \sigma _{S}^{22}$, the
time-independent Hamiltonian is found as%
\begin{equation}
H_{n}^{\prime }=\delta \sigma _{S}^{22}+g\sigma _{A_{1,n}}^{10}\sigma
_{S}^{02}+g\sigma _{A_{2,n}}^{10}\sigma _{S}^{12}+h.c.  \label{T_INDEP}
\end{equation}%
the subscript $\prime $ in the operator $O^{\prime }$ implies that the
operator $O$ is represented in the new rotating frame. After n-th collision
step, the reduced density operator for the system is%
\begin{equation}
\rho _{n}^{\prime }=Tr_{A_{1},A_{2}}\sum_{m,m^{\prime }}e^{-i(\epsilon
_{m}-\epsilon _{m^{\prime }})\tau }\left\langle m\right\vert \sigma
_{n-1}^{\prime }\left\vert m^{\prime }\right\rangle \left\vert
m\right\rangle \left\langle m^{\prime }\right\vert \text{,}  \label{EIGEN}
\end{equation}%
where $\left\vert m\right\rangle $ and $\left\vert m^{\prime }\right\rangle $
are the eigenvectors of the time-independent Hamiltonian $H_{n}^{\prime }$,
with $\epsilon _{m}$ and $\epsilon _{m^{\prime }}$ the corresponding
eigenvalues. Alternatively, $\rho _{n}^{\prime }$ can also be found by
\begin{equation}
\rho _{n}^{\prime }=Tr_{A_{1},A_{2}}\sigma ^{\prime }(\tau ),  \label{LIOU}
\end{equation}%
with $\sigma ^{\prime }(\tau )$ the solution of Liouville equation $\dot{%
\sigma}^{\prime }(t)=-i[H_{n}^{\prime },\sigma ^{\prime }(t)]$ at the time
point $t=\tau $ by considering the initial condition $\sigma ^{\prime
}(0)=\sigma _{n-1}^{\prime }$. Then, the numerical simulation of the system
state after each collision step could be obtained by the numerical
iterations according to Eqn. (\ref{EIGEN}) or (\ref{LIOU}). The numerical
simulation would fit the system's accurate dynamics with high precision
because no approximation has been performed to obtain the analytical form of
Eqns. (\ref{EIGEN}) and (\ref{LIOU}).

\begin{figure}[t]
\centering
\includegraphics[width=8cm]{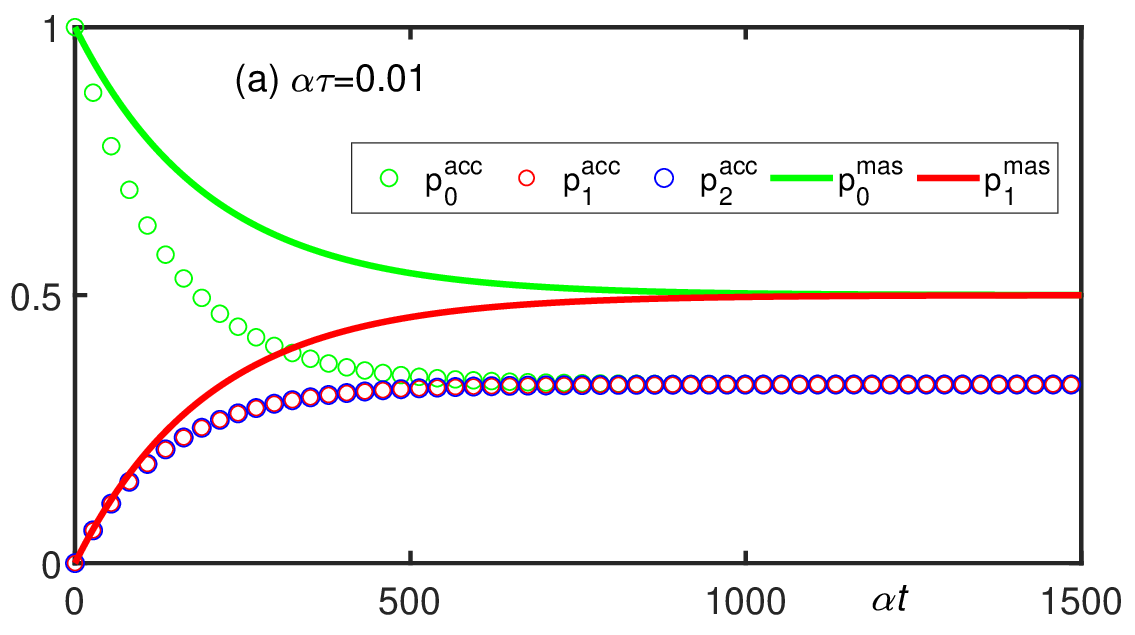}
\includegraphics[width=8cm]{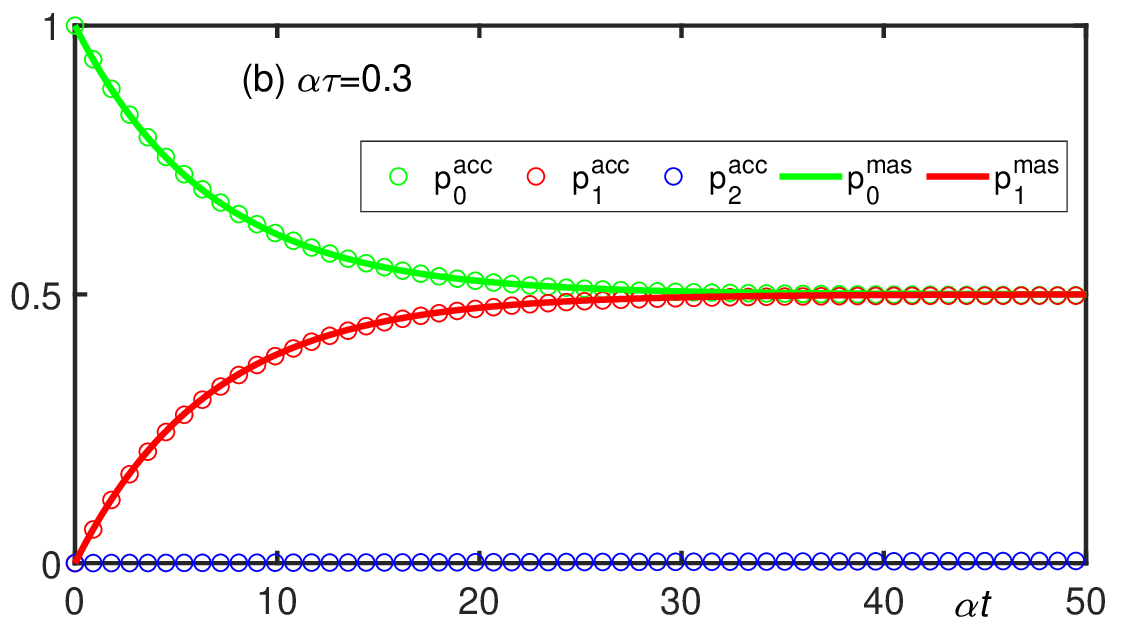}\newline
\caption{Numerical simulations of the collided-system-level-population
evolution governed by the original Hamiltonian (\protect\ref{H_INT}) and the
master equation (\protect\ref{ME}). The system initial state is $%
|0_{S}\rangle $, and the parameters are $\Delta =200g,\protect\omega _{A_{1}}%
\protect\beta _{1}=\protect\omega _{A_{2}}\protect\beta _{2}=10^{-4}$. We
take $\protect\alpha \protect\tau =0.01$ in (a) and $\protect\alpha \protect%
\tau =0.3$ in (b). The original-Hamiltonian-governing populations of levels $%
|0_{S}\rangle $, $|1_{S}\rangle $, and $|2_{S}\rangle $ are labeled by $%
p_{0}^{acc}$, $p_{1}^{acc}$, and $p_{2}^{acc}$, and plotted by the green,
red and blue hollow circles, respectively. The subscript "acc" denotes that
it is the simulation of the system's accurate dynamics. The
master-equation-governing populations of levels $|0_{S}\rangle $ and $%
|1_{S}\rangle $ are labeled by $p_{0}^{mas}$ and $p_{1}^{mas}$, and plotted
by the green and red solid lines, respectively. The blue hollow circles in (a) overlap with the red ones.}
\label{COMP}
\end{figure}

The numerical simulations in Fig. \ref{COMP} compare the accurate dynamics\
of the collided system with the dynamics of the continuous-time master
equation (\ref{ME}) for different collision durations. The system's accurate
dynamics are numerically simulated based on Eqn. (\ref{LIOU}), in which the
numeral solution of Liouville equation\ is derived by the 4-order
Runger-Kutta method. Let us discuss three situations.

In \textit{situation (a), }the collision duration is too small to satisfy
the constraint. The evolution described by Eqn. (\ref{ME}) significantly
differs from the system's accurate evolution, as confirmed in Fig. \ref{COMP}
(a) when $\alpha \tau =0.01$. In the accurate evolution, the collided system
can jump to the level $\left\vert 2_{S}\right\rangle $ with a non-negligible
probability. It can be understood from the fact that, although the
coarse-grained dynamics of the level $\left\vert 2_{S}\right\rangle $ is
negligible in the large detuning case, the level transition between $%
\left\vert 0_{S}\right\rangle $ and $\left\vert 2_{S}\right\rangle $ plays
the predominant role when the evolution duration is in (or smaller than) the
time scale of the rapid-slight-oscillation period, as shown in Fig. \ref%
{VERIFY} (b). In this situation, the coarse-graining on the dynamics does
not work well, and the level $\left\vert 2_{S}\right\rangle $ is not
negligible. The qutrit can not be considered an effective qubit, and hence,
the continuous-time master equation (\ref{ME}) is not valid.

In \textit{situation (b)}, the collision duration is significantly longer
than the time scale of the rapid-slight-oscillation period and much shorter
than the evolution time scale. As shown in Fig. \ref{COMP} (b), the
population evolution governed by Eqn. (\ref{ME}) agrees with the system's
accurate evolution reasonably well. Although the temperatures of the baths
are set significantly large, it is difficult to obtain a noticeable
population of level $\left\vert 2_{S}\right\rangle $. Therefore, the
continuous-time master equation (\ref{ME}) can represent the system dynamics
well.

In \textit{situation (c)}, the collision duration is not short enough
compared to the evolution time scale, so the system dynamics are not
effectively time-continuous. It is challenging to represent the system
dynamics by continuous-time master equations.

\section{Beyond the far-off-resonant case}

According to the dynamics governed by the original Hamiltonian (\ref{H_INT}%
), a continuous-time master equation quite different from the master
equation (\ref{ME}) can be obtained. In the rotating frame introduced in
section \ref{SCD}, the unitary evolution operator is defined as $%
U_{n}^{\prime }=e^{-iH_{n}^{\prime }\tau }$. For the short collision
duration, one can make the approximation up to the second-order of the $\tau
$ and find a continuous-time master equation as%
\begin{eqnarray}
\frac{d\rho ^{\prime }}{dt} &=&i\delta \lbrack \rho ,\sigma _{S}^{22}]+\tau
\delta ^{2}(\sigma _{S}^{22}\rho \sigma _{S}^{22}-\frac{1}{2}[\rho ,\sigma
_{S}^{22}]_{+})  \notag \\
&&+\gamma _{1}e^{\omega _{A_{1}}\beta _{1}}(\sigma _{S}^{02}\rho \sigma
_{S}^{20}-\frac{1}{2}[\rho ,\sigma _{S}^{22}]_{+})  \notag \\
&&+\gamma _{1}(\sigma _{S}^{20}\rho \sigma _{S}^{02}-\frac{1}{2}[\rho
,\sigma _{S}^{00}]_{+})  \notag \\
&&+\gamma _{2}e^{\omega _{A_{2}}\beta _{2}}(\sigma _{S}^{12}\rho \sigma
_{S}^{21}-\frac{1}{2}[\rho ,\sigma _{S}^{22}]_{+})  \notag \\
&&+\gamma _{2}(\sigma _{S}^{21}\rho \sigma _{S}^{12}-\frac{1}{2}[\rho
,\sigma _{S}^{11}]_{+})  \label{ME_A}
\end{eqnarray}%
with $\gamma _{m}=\frac{g^{2}\tau }{1+e^{\omega _{A_{m}}\beta _{m}}}$. It
represents that the qutrit is coupled to bath 1 and bath 2, i.e., the level
transition $\left\vert 0_{S}\right\rangle \leftrightarrow \left\vert
2_{S}\right\rangle $ is coupled to the bath 1, and the transition $%
\left\vert 1_{S}\right\rangle \leftrightarrow \left\vert 2_{S}\right\rangle $
is coupled to the bath 2. In particular, when $\delta =0 $, it turns out to
be the case represented in Ref. \cite{Bera2024}.

Although the continuous-time master equations (\ref{ME}) and (\ref{ME_A})
represent different dynamics, they do not conflict because they work under
different conditions. To represent this in detail, we emphasize that the
approximations up to the second order of $\tau $ are essential to obtain
both Eqns. (\ref{ME}) and (\ref{ME_A}). For Eqn. (\ref{ME}), the
approximation holds when the high-order terms on $\alpha \tau $ can be well
ignored, which can be understood by substituting the expression of $V_{n}$
into the approximate transformation $U_{n}\sim I-iV_{n}\tau -\frac{%
(V_{n}\tau )^{2}}{2}$. Distinctly, for Eqn. (\ref{ME_A}), the approximation
holds when the high-order terms on $\delta \tau $ and $g\tau $ can be well
ignored, which can be understood by substituting the expression of $H_{n}^{\prime }$ into
$U_{n}^{\prime }\sim I-iH_{n}^{\prime }\tau -\frac{(H_{n}^{\prime }\tau )^{2}%
}{2}$. As a consequence, for Eqn. (\ref{ME_A}), it would work well if the
detuning $\delta $ is not extremely large. In contrast, Eqn. (\ref{ME}),
stemming from the effective Hamiltonian (\ref{H_EFF}), works well under
the large-detuning case. Consequently, the larger the detuning $\delta $ is, the
better the performance of Eqn. (\ref{ME}) is. Besides, one can verify that,
by the parameters in Fig. \ref{COMP}, it is challenging to perform the
approximations up to the second-order of $\tau $ to obtain Eqn. (\ref{ME_A})
when the collision duration is significantly longer than the time scale of
the rapid-slight-oscillation period.

\section{conclusions}

In conclusion, we propose a quantum collision model in which an open system
composed of a qutrit off-resonantly collides with the ancillas of two baths.
The qutrit interacts with two ancillas from two baths in each collision
step. In the far-off-resonant case, according to a legal adiabatic
elimination on the highest level of the qutrit, we show that the
qutrit-ancilla interaction is effectively considered as the two ancillas
commonly driving the level transition of a qubit. A master equation, which
implies that our scheme is considered a qubit coupled to an effective bath,
is found in the continuous-time limit via coarse-graining. It shows that the
temperature of the effective bath can be an arbitrary negative real number.
The validity of the approach is confirmed by numerically comparing the
system's accurate dynamics to the qubit dynamics represented by the master
equation. Because the far-off-resonant interaction results in rapid
oscillation dynamics, the collision duration should be significantly longer
than the rapid oscillation period to ensure that each collision process is
coarse-grained. It differs from the quantum collision model working in a
resonant regime as there are no rapid oscillation dynamics. Beyond the
far-off-resonant case, the system dynamics can be represented by a master
equation implying a qutrit coupled to two independent baths, where the
highest level of the qutrit can not be legally adiabatically eliminated, and
the qutrit is not regarded as an effective qubit. The work outlines an
approach to couple a qubit with an effective negative temperature based on
quantum collision model. It systematically investigates the quantum
collision model in the off-resonant regime. It also implies the approaches
to realize an effective negative temperature coupled to a qubit based on
conventional thermal baths. For example, a qutrit is coupled to two
conventional thermal baths, where the frequencies of the baths are filtered
so that they are far different from the corresponding level transition
frequencies of the qutrit. Alternatively, it implies a theoretically
reasonable but experimentally challenging approach in which two conventional
thermal baths are coupled to a common qubit.



\end{document}